\documentstyle{mn}

\begin{document}
\label{firstpage}

\def\msun{{\rm M}_\odot}
\def\rsun{{\rm R}_\odot}

\title{Type Ia Supernovae and Remnant Neutron Stars}
\author[A. R.~King, J.E.~Pringle \& D.T.~Wickramasinghe]{A. R.~King$^{1, 2}$,
J.E.~Pringle$^{2}$, D.T.~Wickramasinghe$^{2, 3}$ \\
   $^1$ Astronomy Group, University of Leicester,
        Leicester, LE1~7RH\\
   $^2$ Institute of Astronomy, Madingley Rd, 
        Cambridge CB3~OHA\\
   $^3$ Astrophysical Theory Centre, Australian National University, ACT 0200,
                  Australia}
\date{Accepted. Received}
\pagerange{\pageref{firstpage}--\pageref{lastpage}}
\pubyear{2000}

%-----------less/greater than approx eq to--------------
\newcommand{\lta}{{\small\raisebox{-0.6ex}{$\,\stackrel
{\raisebox{-.2ex}{$\textstyle <$}}{\sim}\,$}}}
\newcommand{\gta}{{\small\raisebox{-0.6ex}{$\,\stackrel
{\raisebox{-.2ex}{$\textstyle >$}}{\sim}\,$}}}   
%-------------------------------------------------------

\maketitle
\begin{abstract}
On the basis of the current observational evidence, we put forward the
case that the merger of two CO white dwarfs produces both a Type Ia
supernova explosion and a stellar remnant, the latter in the form of a
magnetar. The estimated occurrence rates raise the possibility that
many, if not most, SNe Ia might result from white dwarf mergers.
\end{abstract}

\section{Introduction}
\label{sec:intro}

Type Ia supernovae are currently being used as empirical standard
candles in the redshift range $z = 0.1 - 1$ to provide evidence that
the expansion of the Universe might be accelerating (Riess et al.,
1998; Perlmutter et al., 1998, 1999). This has naturally renewed
interest in what they might be. The case for cosmological acceleration
depends critically on the degree to which Type Ia supernovae can be
treated as standard candles, independent of redshift. Only a good
understanding of the nature and the origin of these supernovae
can provide confidence that they actually are standard candles.

The most favoured scenario (see, for example, Livio 2000) for the SN
Ia event is the explosion and complete disintegration of a CO white
dwarf, brought about by the accretion of material which pushes the
mass of the white dwarf over the Chandrasekhar limiting mass $M_{\rm
Ch} = 1.44 \msun$. The resulting rapid conversion of about a solar
mass of C/O to Ni$^{56}$, and the subsequent decay of Ni$^{56}$ to
Fe$^{56}$, provides the right amount of energy to power the observed
explosion, and releases it on the right timescale to explain the
observed light-curve. It can also account for the lack of hydrogen
observed in these supernovae.

While this picture is largely agreed, the main debate concerns the
nature of the precursor driving the accretion which pushes the white
dwarf mass over the limit. Here the favoured view envisages two main
possibilities.

The first possibility is that the precursor is a binary system
containing a white dwarf accreting hydrogen from a non--degenerate star
(the single--degenerate scenario). This possibility suffers from two
main drawbacks. First, the accretion of hydrogen on to a white dwarf
can lead to ordinary nova explosions, which over time tend to decrease,
rather than increase, the mass of the white dwarf; and second, it is
difficult, though not perhaps impossible, to set off a supernova
explosion right next to a large mass of hydrogen (the non-degenerate
companion) without the supernova ejecta becoming contaminated
by the hydrogen from the companion's envelope (Marietta, Fryxell \&
Burrows, 2000).

The second possibility is the merger of two white dwarfs (the
double--degenerate scenario). The white dwarfs are brought together by
gravitational radiation on a timescale $t_{\rm grav}$, until the less
massive, and thus the less dense, fills its Roche lobe and begins to
transfer mass. The initial timescale for mass transfer is set by
gravitational radiation, and for these systems is of order $10^6$
years. Many authors (Saio \& Nomoto, 1985, 1998; Kawai, Saio \&
Nomoto, 1987; Mochkovitch \& Livio, 1990; Timmes, Woosley \& Taam,
1994; Mochkovitch, Guerrero \& Segretain, 1997) contend that under
such circumstances no explosion takes place, and the result is a
quiet, accretion--induced collapse (AIC), forming a neutron star
remnant.

In this paper we argue that there is strong observational evidence
that the merger of two CO white dwarfs produces {\it both} a supernova
explosion {\it and} a stellar remnant; and further, that since such a
supernova does not involve hydrogen it must be of Type I, and probably
of Type Ia. Thus at least some, if not all, SNe Ia result from the
merging of two white dwarfs. In Section 2, we consider the outcome of
the merger if the total mass of the two white dwarfs is less than
$M_{\rm Ch}$, and identify the likely merger products as massive and
highly magnetic white dwarfs.  In Section 3 we consider the case where
the combined mass exceeds $M_{\rm Ch}$, and by analogy identify the
likely merger products as the magnetars. We summarize our conclusions
in Section 4.

\section{Merger Products}

Since the outcome of a CO white dwarf merger is difficult to predict
theoretically, we start from a case where the answer is clear, namely
when the total binary mass $M$ is slightly smaller than $M_{\rm
Ch}$. A supernova is unlikely (but see Section 3 below), so there must
be a remnant -- a massive white dwarf. Indeed Livio, Pringle \& Saffer
(1992) suggested that a significant fraction of massive white dwarfs
are the result of mergers. Furthermore the white dwarf is spun up to
rapid rotation by accretion from a disc, and is likely to be highly
magnetic because of the winding up of magnetic fields in this
disc. Statistical evidence supports this picture. It is now well
established that the mass distribution of isolated white dwarfs has,
in addition to the dominant peak at 0.57 $\msun$, a second peak near
1.2 $\msun$ with a tail which extends up to $M_{\rm
Ch}$. Wickramasinghe \& Ferrario (2000) show that a large proportion
(about 25 per cent) of the white dwarfs in this high mass group are
strongly magnetic, while for the white dwarf sample as a whole, only 5
per cent are magnetic. However it is unlikely that {\it all} high mass
magnetic white dwarfs result from mergers, since some rotate very
slowly (periods $>100$~yr). These must arise from single star
evolution (see Wickramasinghe and Ferrario 2000).

We should next ask for a specific example of such a merger remnant.
The best studied massive magnetic white dwarf is RE~J0317--853, which
has mass $M_{\rm WD} =1.35\msun$, magnetic field $B_0$ in the range
$3.5\times 10^8 - 8\times 10^8$~G, and spin period $P_0 = 725$~s (see
Wickramasinghe \& Ferrario 2000 and references therein). This looks
remarkably like a white dwarf merger product which missed $M_{\rm Ch}$
by a narrow margin. However before accepting this important conclusion
we should examine other possibilities.

\subsection{Single--star evolution}

RE~J0317--853 could in principle have formed in the normal course of
single--star evolution as the degenerate core of a giant. Its mass
$M_{\rm WD} \la M_{\rm Ch}$ implies that the latter star must have had
a mass close to the maximum that will give a white dwarf rather than a
neutron star or black hole, i.e. about $8\msun$. Livio \& Pringle
(1998) argue that dynamo--generated magnetic fields at the
core--envelope interface will make the core of such a star rotate with
angular velocity $\Omega_c \simeq 6.6\times 10^{-11}$~s$^{-1}$ at the
end of the giant phase. The inner $1.35\msun$ of this core has radius
$0.2\rsun$: collapsing this to the likely radius $R_0 \sim 3\times
10^8$~cm of RE~J0317--853 and assuming angular momentum conservation
produces a spin period of about $4\times 10^7$~s.  This is probably an
underestimate, as the white dwarf magnetic field implied by Livio \&
Pringle's calculations is much smaller than the observed $B_0 =
3.5-8\times 10^8$~G: the core field at the end of the giant phase is
$B_c \simeq 2\times 10^{-2}$~G, and flux conservation increases this
only to $\sim 50$~G for the white dwarf. Spruit \& Phinney (1998)
predict somewhat longer white dwarf spin periods $\sim 10^8$~s, as in
their calculations the degenerate core is close to corotation with the
giant envelope. We conclude that RE~J0317--853's observed spin period
$P_0 = 725$~s cannot be explained if it is the result of single--star
evolution. Further, this evolution offers no obvious reason why the
observed magnetic field should be so strong.

\subsection{Binary evolution}

Descent from an interacting binary offers a clear avenue for
explaining the rapid spin of RE~J0317--853.

(a) {\it Conventional CV evolution.}

The most straightforward idea is that RE~J0317--853 might represent
some endpoint of cataclysmic variable (CV) evolution, in which a white
dwarf accretes from a low--mass companion.  RE~J0317--853's strong
field would make it an extreme member of the AM Herculis subgroup (in
fact its field is stronger than any known member of this class). In
the conventional picture of AM Herculis evolution, the strong field of
the white dwarf keeps the spin of this star locked to the orbital
motion. RE~J0317--853 cannot descend from this evolution, as the
minimum orbital period for any CV is about 80 minutes, far above the
observed $P_0 = 725$~s.

(b) {\it Unusual CV evolution}
 
In a recent paper, Meyer \& Meyer--Hofmeister (1999) argue that AM Her
systems may lose synchronism at very short orbital periods, when the
secondary becomes so cool that the conductivity of its envelope drops
catastrophically. In this case the white dwarf could indeed spin up to
much shorter periods, and the companion star would be disrupted on a
short timescale as the binary separation shrinks because of the
draining of orbital angular momentum to the white dwarf spin.  At
first sight this looks like an attractive idea for explaining the
properties of RE~J0317--853. However since the deeper layers of the
companion must remain ionized, we would expect this star to retain a
strong enough dipole moment to remain synchronous. Even leaving this
aside, this idea offers no explanations for the unusually high mass
and magnetic field of RE~J0317--853.

We conclude that RE~J0317--853 is not likely to be be explained as an
end--product of these other types of evolution. On the other hand, as
we suggested above, it arises quite naturally as the result of a CO
white dwarf merger, with at least one of the white dwarfs being mildly
magnetic. A variant of this idea is to invoke coalescence of such a CO
white dwarf with the fairly massive core of a giant companion through
common--envelope evolution. For many purposes these two possibilities
are extremely similar.

\section{What if $M > M_{\rm Ch}$?}

We concluded above that the massive white dwarf RE~J0317--853 is the
result of a white dwarf merger with $M$ slightly less than $M_{\rm
Ch}$.  We can now ask what end--product would have emerged had $M$
been slightly larger then $M_{\rm Ch}$, and the resulting collapse had
left a remnant rather than provoking complete disruption.

We first consider the merger process in a little more detail.  If the
mass ratio $q$ is less than 0.63 mass transfer is stable, and
continues at a rate governed by gravitational radiation. However, if
$q > 0.63$, mass transfer is dynamically unstable. Mass is then
transferred rapidly until $q < 0.63$. Stability is achieved on a
timescale of $ \tau \sim t_{\rm grav}^{2/3}P^{1/3}$, where $P$ is the
orbital period and $t_{\rm grav}$ is the timescale specifying the rate
at which transfer begins. Typically we expect $t_{\rm grav} \sim
10^6$~yr, $P\sim$ a few hours, and thus $\tau$ to be of order a few
hundred years. However, the mass transfer rate for an $n=3/2$
polytrope obeys $\dot{M}/M \sim P^{1/2}(t_0 - t)^{-3/2}$, valid for
time $t$ less than some reference time $t_0$, (Webbink 1985), so the
bulk of the mass transfer before stability is achieved occurs on a
timescale of several orbital periods. Once stability is achieved,
transfer slows once more towards the rate governed by gravitational
radiation.

Because no existing computation has been able to consider accretion of
He or of C/O on to a white dwarf at such high rates, there is still
considerable uncertainty as to what the final outcome might be. For
example, Reg\H os et al (2000) argue, from population synthesis
models, that the majority of SNe Ia are caused by rapid accretion of
He on to a sub--Chandrasekhar--mass white dwarf and a subsequent
edge--lit detonation of carbon, leading to the complete thermonuclear
disintegration of the white dwarf. In contrast, as we remarked above,
many authors contend that such edge--lit ignition can lead to quiet
burning of the CO to O/Ne/Mg, and thus speculatively to a quiet
accretion induced collapse (at least for $M = M_{\rm Ch}$) to form a
neutron star but with no supernova explosion and thus with no
supernova remnant. While the computations have yet to be carried out,
it seems to us hard to escape the conclusion that if the Chandrasekhar
limit is exceeded during the mass transfer, collapse to neutron star
densities must ensue.

If, during such a collapse, we assume conservation of angular momentum
and magnetic flux as the stellar radius shrinks from the $R_0 \simeq
3\times 10^8$~cm of RE~J0317--853 to the $R= 10^6$~cm of a neutron
star, we find a spin period $P = P_0(R/R_0)^2 = 7$~ms and a field $B =
B_0(R_0/R)^2 = 3.5 - 8\times 10^{13}$~G. We draw attention to the fact
that these values are remarkably close to those required in the
magnetar model now thought to provide an explanation of the properties
of soft gamma repeaters (SGRs) and the related anomalous X--ray
pulsars (AXPs) (see Thompson, 1999, and Kouveliotou, 1999 for recent
reviews). Moreover, we might expect even more extreme values of these
two parameters for two reasons. First, the strongly increased shearing
resulting from a collapse to much smaller dimensions is likely to
increase the strength of the magnetic field considerably (Thompson \&
Duncan, 1995; Kluzniak \& Ruderman, 1998). Second, with a surface
temperature of $\sim 4\times 10^4$~K, RE~J0317--853's spin
period at birth could have been considerably shorter than
the current 725~s, as even tiny amounts of mass loss coupling to its
large magnetic moment would have caused spindown within its cooling
age of several $10^7$~yr. 

More extreme fields and rotation rates put us squarely in the
parameter space ($P = $ few ms, $B \ga 10^{14}$~G) inferred for
magnetars at birth. We conclude that the probable outcome of a
magnetic white dwarf merger with $M > M_{\rm Ch}$ is a magnetar.

\section{Discussion}

We have argued that the merger of two CO white dwarfs results in a
remnant which is both rapidly rotating and highly magnetic. If the
total mass is less than $M_{\rm Ch}$, the remnant is a massive,
magnetic white dwarf. And if the total mass exceeds $M_{\rm Ch}$, the
remnant is a rapidly rotating, strongly magnetic neutron star. We have
identified such remnants as magnetars.

This conclusion leads to another. Soft gamma ray repeaters (and
anomalous X--ray pulsars) are associated with supernova remnants (see,
for example, the discussion in Kouveliotou, 1999). This implies that
the collapse caused by the merger is not a quiescent `accretion
induced collapse', but actually gives rise to a supernova
explosion. If our identification of magnetars as CO--CO white dwarf
merger products is correct, then the supernovae associated
with them should have high--velocity carbon and higher--mass elements
(from the disrupted remnant disc), but no hydrogen or helium. 
We conclude, therefore, that CO--CO white dwarf mergers produce both a
Type I supernova and a neutron star remnant, and, further, that if one
of the merging white dwarfs has a significant magnetic field
(estimated at around 25 per cent of the total), then this neutron star
is a magnetar. 

We may use this now to estimate an occurrence rate for these
supernovae. SGRs and AXPs are known to have rather short lifetimes
$\sim 10^4$~yr, (cf Kouveliotou, 1999; Thompson, 1999), from arguments
based on the observed spindown timescale $|P/\dot P|$, and the typical
age of the associated supernova remnants. The magnetar model gives
similar (or even shorter) spindown ages. From the current observed
total number of SGRs and AXPs ($\sim 10$), this characteristic age
implies an estimate for the formation rate of magnetars of $\sim
10^{-3}$~yr$^{-1}$ in the Galaxy. 

We now ask: what kind of Type I supernovae do the CO--CO--mergers
correspond to?  We note first that the inferred SNe Ia rate for the
Galaxy is approximately $\sim 10^{-3}$ yr$^{-1}$ (Yungelson \& Livio
2000). Thus the estimated occurrence rates provide no obvious grounds
for rejecting the possibility that {\it most} SNe Ia might result from
white dwarf mergers. Moreover they cannot be of Type Ib, which are
associated with high-mass stars, and, if the above estimates are
correct, they are too numerous to be of Type Ic (Cappellaro et al.,
1997 show that the combined rate for Types Ib and Ic is lower than for
Type Ia).

The white dwarf merger scenario is currently perhaps the less favoured
option for SNe Ia, but there are not yet adequate grounds to rule it
out. While the weight of opinion appears to be that the merger of two
white dwarfs leads to accretion-induced collapse, and no supernova
explosion or remnant, the computations required to provide
verification have yet to be carried out. Moreover, Reg\H os et al
(2000), on the basis of population synthesis calculations, conclude
that most SNe Ia result from edge--lit detonations in merging white
dwarfs with $M < M_{\rm Ch}$. Here we have argued that merging white
dwarfs with $M > M_{\rm Ch}$ might also give rise to SNe Ia. In both
cases, the homogeneity of the initial conditions, and the available
energy supply, point to uniformity of outcome which is characteristic
of SNe Ia. And in neither case have the necessary computations been
performed to determine what that outcome might be.

\section{Acknowledgments}
We thank Carole Haswell, Chryssa Kouveliotou, Chris Thompson and Chris
Tout for helpful conversations, and David Branch for a helpful
referee's report. ARK gratefully acknowledges a PPARC Senior
Fellowship, and he and DTW thank the Institute of Astronomy,
Cambridge, for hospitality.

\label{lastpage}
\end{document}